\documentclass[%
reprint,
amsmath,
amssymb,
aps,
pra,
superscriptaddress,
preprintnumbers,
prstab,
prstper]{revtex4-2}
\usepackage{graphicx}
\usepackage{dcolumn}
\usepackage{bm}

\usepackage[english]{babel}
\usepackage[utf8]{inputenc}
\usepackage{amssymb,amsfonts,amsmath,mathtools,mathrsfs}
\usepackage{relsize}
\usepackage{mdframed}
\usepackage{enumitem}
\usepackage{graphicx}
\usepackage{url,hyperref}
\usepackage[table,usenames,dvipsnames]{xcolor}
\usepackage{gensymb}
\usepackage{grffile}
\usepackage{float}

\definecolor{darkgreen}{rgb}{0.0, 0.5, 0.0}

\hypersetup{
colorlinks=true,
linkcolor = blue,
filecolor = blue,
urlcolor = Blue,
citecolor = blue,
pdftitle = {Peierls Rashba Dresselhaus}
}

\usepackage[capitalize]{cleveref}
\usepackage{textcomp}
\usepackage{braket} 

\begin{document}
\title{Stabilization of Biskyrmions by chiral interaction in centrosymmetric magnets }

		\author{Deepak S. Kathyat}
		\affiliation{School of Physical and Mathematical Sciences, Nanyang Technological University, Singapore}
		\author{Pinaki Sengupta}
		\affiliation{School of Physical and Mathematical Sciences, Nanyang Technological University, Singapore}

\begin{abstract}
We report a microscopic mechanism for stabilization of biskyrmions in nature by investigating a minimal classical spin lattice model with nearest neighbour ferromagnetic Heisenberg  exchange and static chiral magnetic interaction on the triangular lattice. The model is physically motivated model from the Mott insulators with broken time-reversal symmetry, that is, in the large-$U$ limit of the one band Hubbard model at half-filling on a two-dimensional lattice in the presence of external magnetic field. At order $1/U^{2}$, the external magnetic field can induce a chiral interaction between the three neighbouring spins. We demonstrate that the chiral magnetic interaction results in biskyrmion states above a critical value and its strength affects the size of biskyrmions forming.
\end{abstract}
\date{\today}

\maketitle

\noindent
\underline{\it Introduction:} Topologically protected magnetic textures like skyrmions and antiskyrmions can enable spintronic applications for  building information storage and processing devices in future \cite{Fert2017, Wiesendanger2016, Fert2013, Nagaosa2013b, Bogdanov2020, Gobel2019, Karube2018}. Till date such textures are observed in wide range of materials in various experimental studies \cite{Dupe2014, Pollard2017, Soumyanarayanan2017, Romming2013b, Yu2012, Yu2011, Zhao2016, Meyer2019, Tonomura2012, Hirschberger2019,Jin2017, Muhlbauer2009b,Yu2018, Yu2010a, Hoffmann2017, Nayak2017,niitsu2022geometrically}. In centrosymmetric materials the absence of inversion symmetry breaking forbids Dzyaloshinskii-Moriya interaction (DMI). In these materials the skyrmion crystals gets stabilized by competeing exchange interactions coming from geometrical frustration of the short-range two-spin interactions hence termed as ‘chiral geometric frustration’ \cite{PhysRevLett.108.017206,batista2016frustration,lin2016ginzburg,leonov2015multiply} or magnetic anisotropies \cite{PhysRevB.103.104408}. In non-centrosymmetric crystals, formation of these textures is understood theoretically via interplay between Heisenberg exchange and DMI found in materials lacking inversion symmetry \cite{farrell2014,chen2016,roessler,mohanta2019,chen2016,rossler2010,iwasaki2014,Yi2009}. In metallic magnets with non-centrosymmetric geometry, the relevant microscopic models feature spin-orbit coupling of itinerant electrons \cite{Hayami2018,Hayami2019,Bezvershenko2018,hayami2021field,jin2021field,Kathyat2020a,Kathyat2021}. Recent studies have elucidated the connection between topological magnetic textures and the underlying electronic band structure in metallic magnets \cite{kathyat2021antiskyrmions,mukherjee2021engineering}. Such magnetic textures can be controlled using ultra-low currents in metals. In contrast, complex magnetic textures affect the electronic properties by inducing emergent magnetic fields acting on conduction electrons through imparting
a Berry phase and lead to magneto-electric phenomena such as the colossal magnetoresistance and the topological Hall effect \cite{Kathyat2021}. At very low current densities these magnetic textures lead to large spin-transfer torques, which has strong implications for spintronic applications..

\noindent 
Topological magnetic textures are described by an integer, known as topological charge or topological number, $n_{sk}$ which is defined as the integral of the solid angles spanned by the three neighbouring spins and is given by \cite{Kovalev2018Sep}:
\begin{equation}
\label{eq:Tcharge}
n_{sk}=\frac{1}{4\pi}\int d^2r(\partial_x \textbf{m}\times \partial_y\textbf{m})\cdot\textbf{m}
\end{equation}
where, \textbf{m} is a unit vector pointing in the direction of the magnetization. The topological charge describes how many times magnetic moments wrap around a unit sphere upon application of stereographic
projection. As a consequence, skyrmions possess topological protection, since erasing a skyrmion requires globally modifying the system - local changes in spin configurations are not sufficient. Skyrmions and antiskyrmions have opposite topological charge of $\pm 1$. Skyrmions with higher $n_{sk}$ can bring about richer physics. For example, a large topological Hall effect can be observed because of the larger spin noncoplanarity which can induce larger emergent electro-magnetic fields for the conduction electrons \cite{PhysRevLett.118.147205}. They may also allow multiple digital control of the topological numbers.

Biskyrmions, the magnetic textures of $n_{sk} = \pm 2$ are composed of two bound skyrmions of oppositely swirling spins (magnetic helicities) are rarely found. The first real space observation of biskyrmions was shown using Lorentz Transmission electron Microscopy (LTEM) in 2014 in a thin plate of bilayer manganite $\text{La}_{2-2x}\text{Sr}_{1+2x}\text{Mn}_{2}\text{O}_{7}$ with uniaxial magnetic anisotropy \cite{Yu2014Jan}. It was also demonstrated that biskyrmions can be electrically driven with much lower current density as compared to the conventional ferromagnetic domain walls. Later, these textures were observed in other centrosymmetric magnets such as $\text{MnNiGa}$, $\text{Nd}_{2}\text{Co}_{17}$ \cite{Peng2017Nov,Zuo2023Jan}. Although there is an ongoing debate on the observations of biskyrmion crystals by LTEM imaging. In two recent publications, the authors showed that the tilting effect can also be responsible for the unique Lorentz contrast. Tubes of topologically-trivial bubbles appear as skyrmion pairs with reversed in-plane magnetizations when viewed under an angle \cite{Loudon2019Apr,Yao2019Mar}. Several theories have been put forward to explain the stabilization mechanism of biskyrmions \cite{Gobel2019Jul,PhysRevB.100.014432,PhysRevResearch.1.033011,PhysRevLett.118.147205,PhysRevLett.129.017201,Wang_2022}. Gobel, et.al. postulate stabilization of biskyrnmions by dipole-dipole interacion. This is present in all magnetic materials, but gets overshadowed by DMI in non centrosymmetric materials. In centrosymmetric materials, a short range dipole-dipole interaction leads to attraction of two skyrmions resulting in formation of biskyrmions which is energetically preferable over two individual skyrmions\cite{Gobel2019Jul}. Biskyrmions arranged in a triangular lattice are found to exist as a result of the competing nearest-neighbour exchange interaction, perpendicular magnetic anisotropy, dipole-dipole interaction, and the applied external magnetic field \cite{PhysRevB.100.014432,PhysRevResearch.1.033011}. In addition to insulating magnets, biskyrmions are also found to be stable in itinerant magnets at zero magnetic field. The Ruderman–Kittel–Kasuya–Yosida (RKKY) interaction between the localized magnetic moments mediated by conduction electrons plays crucial role in stabilizing biskyrmions as shown using modified kernel polynomial method with Langevin dynamics (KPM-LD) simulations\cite{PhysRevLett.118.147205,PhysRevLett.129.017201}. Formation of biskyrmions mediated by an intrinsic emergent monopole-antimonopole pair has been demonstrated using micromagnetic simulations \cite{Wang_2022}.

In this work, we propose a simple chiral spin interaction model resulting from the large-U limit of the
one band Hubbard model at half-filling in the presence of external magnetic field and show the emergence of biskyrmions on centrosymmetric triangular lattice. Our focus is on the magnetic ground states; thermally stabilized magnetic textures are deferred to later studies. The appearance of biskyrmions is explicitly demonstrated using the state-of-the-art classical Monte Carlo simulations. We show that the competeing nearest-neighbour exchange interaction and scalar chiral magnetic interaction give rise to emergence of biskyrmions.

\noindent
\underline{\it Model and Method:} We start with an effective spin model on triangular lattice\cite{PhysRevB.51.1922,bauer2014chiral},


\begin{eqnarray}
H & = & J_H \sum_{\langle i,j \rangle} {\bf S}_i \cdot {\bf S}_j + J_C \sum_{i,j,k \in \triangle } {\bf S}_i \cdot  ({\bf S}_j \times {\bf S}_k) .
\label{eq:chiral}
\end{eqnarray} 
The first term in the Hamiltonian Eq. (\ref{eq:chiral}) is the Heisenberg exchange interaction with coupling, $J_H$ between spins at nearest neighbor sites $\langle ij \rangle$. The second term is chiral magnetic interaction with coupling strength, $J_C$. $i,j,k$ are sites in counter-clockwise direction around each elementary triangle. This term is often studied to understand chiral topological spin liquid phase in systems with quantum spins \cite{samajdar2019enhanced,bauer2014chiral,PhysRevB.94.075131,PhysRevB.92.125122}. Systems with ions that carry a large moment, the localized spins can be approximated as classical vectors. Now we study the Hamiltonian (Eq. (\ref{eq:chiral})), treating the spins as classical vectors.

\begin{figure}[h]
	\centering
	\includegraphics[width=1.0 \columnwidth,angle=0,clip=true]{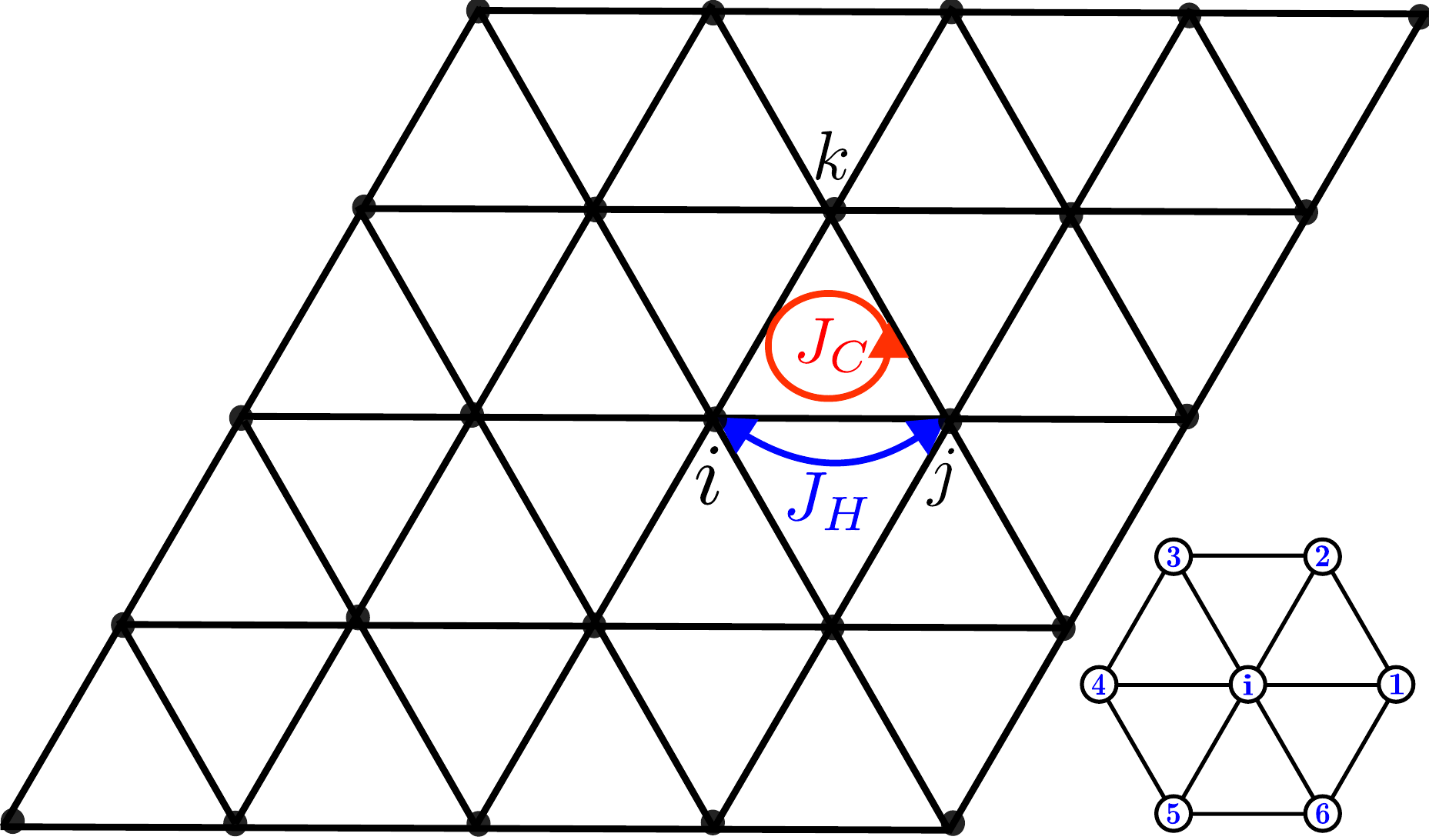}
	\caption{(Color online) Schematic diagram showing various terms in the Hamiltonian. In a triangular lattice $J_{H}$ is nearest neighbour ferromagnetic exchange interaction between sites $i$ and $j$. $J_{C}$ is the chiral magnetic interaction for each elementary triangle of the lattice with sites $i$, $j$ and $k$ in anti-clock wise direction. Right bottom panel shows the sites involved in the calculation of local skyrmion density $\mathcal{T}_{i}$ at site $i$.
	}
	\label{fig:triangle}
\end{figure}

We compute the component resolved spin structure factor (SSF) to characterize the ordered magnetic phases. The components of SSF is given by,
\begin{eqnarray}
 S^{\mu}_{f}({\bf q}) &=& \frac{1}{N^2} \bigg \langle \sum_{ij} S^{\mu}_i S^{\mu}_j~ e^{-{\rm i}{\bf q} \cdot ({\bf r}_i - {\bf r}_j)} \bigg \rangle
\label{SSF}
\end{eqnarray}
\noindent
with $\mu = x,y,z$. The total SSF is $S_f({\bf q}) = S^{x}_f({\bf q}) + S^{y}_f({\bf q}) + S^{z}_f({\bf q})$. The perpendicular and parallel components of SSF are $S^{\perp}_f({\bf q}) = S^{z}_f({\bf q})$, $S^{\parallel}_f({\bf q}) = S^{x}_f({\bf q}) + S^{y}_f({\bf q})$ respectively. 

We also calculate the discretized version of topological charge called as skyrmion density, $\mathcal{T}$, given by \cite{PhysRevB.92.214439},
\begin{equation}
 \centering
 \mathcal{T} = \frac{1}{4\pi} \Bigg \langle  \sum_i A_i^{(12)} \text{sgn} [\mathcal{L}^{(12)}_{i}] + A_i^{(45)} \text{sgn} [\mathcal{L}^{(45)}_{i}]  \Bigg \rangle,
\end{equation}
which is total of the local skyrmion densities, $ \mathcal{T}_{i}$ at each lattice site.
\begin{equation}
 \centering
 \mathcal{L} = \frac{1}{8\pi} \Bigg \langle  \sum_i \mathcal{L}_i^{(12)} + \mathcal{L}_i^{(45)} \Bigg \rangle,
\end{equation} 
\noindent
where, $A_i^{(ab)} = ||({\bf S}_{i_a} - {\bf S}_{i}) \times ({\bf S}_{i_b} - {\bf S}_{i})||/2$ is the local area of the surface spanned by three spins on every elementary triangular plaquette ${\bf r}_i,{\bf r}_a,{\bf r}_b$. Here $\mathcal{L}_{i}^{(ab)} = {\bf S}_i.({\bf S}_{i_a} \times {\bf S}_{i_b})$ is the so-called local chirality and ${\bf r}_i, {\bf r}_1 - {\bf r}_5$ (see right bottom panel Fig. \ref{fig:triangle}) are the sites involved in the calculation of $\mathcal{T}$. 

\noindent

\noindent
\underline{\it Results:} 
\begin{figure*}[t!] 
\centering
	\includegraphics[width=2.0 \columnwidth,angle=0,clip=true]{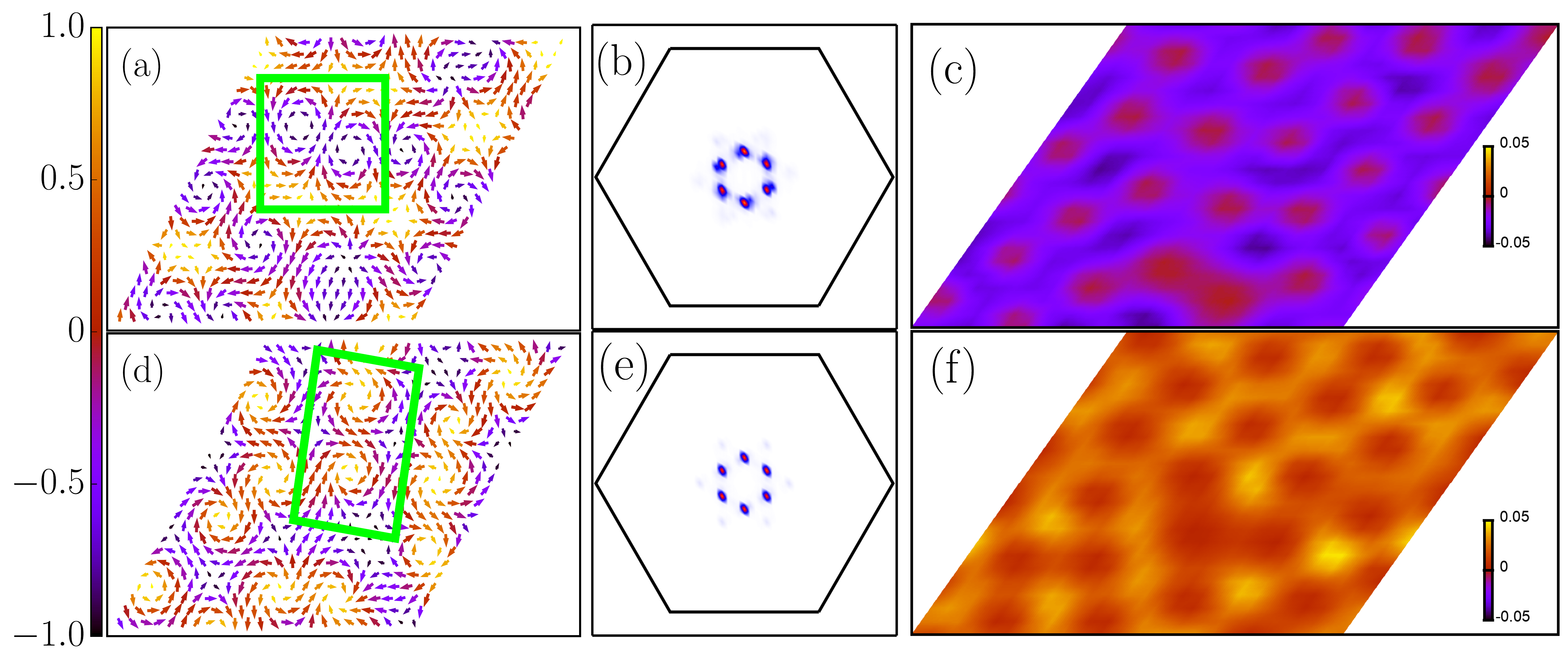}
	\caption{(Color online) Ground state spin configurations at low temperature ($T = 0.001$ ) for (a) $J_C = -1.0$ and (d) $J_C = 1.0$ on triangular lattice of size $N = 24^2$. The green box highlights the biskyrmions in both cases. (b) and (e) are the spin structure factor plots corresponding to magnetic states (a) and (d) respectively. (c) and (f) shows the corresponding local skyrmion density $\mathcal{T}_{i}$ maps on the lattice.
	}
	\label{fig:spin}
\end{figure*}

The first term in the Hamiltonian Eq. (\ref{eq:chiral}) is Heisenberg model which has ferromagnetic ground state for $J_H = -1.0$. We investigate the effect of chiral magnetic interaction on the ferromagnetic exchange Heisenberg model. We use classical Monte Carlo simulations with the standard Metropolis algorithm. The simulations are carried out on lattice sizes varying from $N = 40^2$ to $N = 100^2$, and $\sim 5\times 10^4$ Monte Carlo steps are used for equilibration and averaging at each temperature point. Spin configurations in the magnetic states obtained at low temperature ($T = 0.001$ ) are shown for lattice size $N = 24^2$ for $J_C = -1.0$ in (see \cref{fig:spin}($a$)) and $J_C = 1.0$ in (see \cref{fig:spin}($d$)). We can clearly see the stabilization of Bloch biskyrmions in both cases. The central spins of biskyrmions in the (\cref{fig:spin}($a$)) are pointing down while the biskyrmions in the (\cref{fig:spin}($d$)) have central spins pointing up. The spin structure factor for both the magnetic configurations  (see \cref{fig:spin}($b$ and $e$)) shows the hexagonal peaks. The quatintity that differentiate these two types of biskyrmions is their skyrmion density, $\mathcal{T}$. Local skyrmion density maps on the lattice are shown in (\cref{fig:spin}($c$ and $f$)) corresponding to the magnetic textures in (\cref{fig:spin}($a$ and $d$)) respectively. An ideal biskyrmion of type  (\cref{fig:spin}($a$)) has $n_{sk} = -2$ as the central spin is pointing down. We can see the corresponding skyrmion densities are negative on the local skyrmion density map. Similarly in (\cref{fig:spin}($f$)), local skyrmion densities are positive for the biskyrmions with $n_{sk} = +2$.

\begin{figure}[H]
	\centering
	\includegraphics[width=1.0 \columnwidth,angle=0,clip=true]{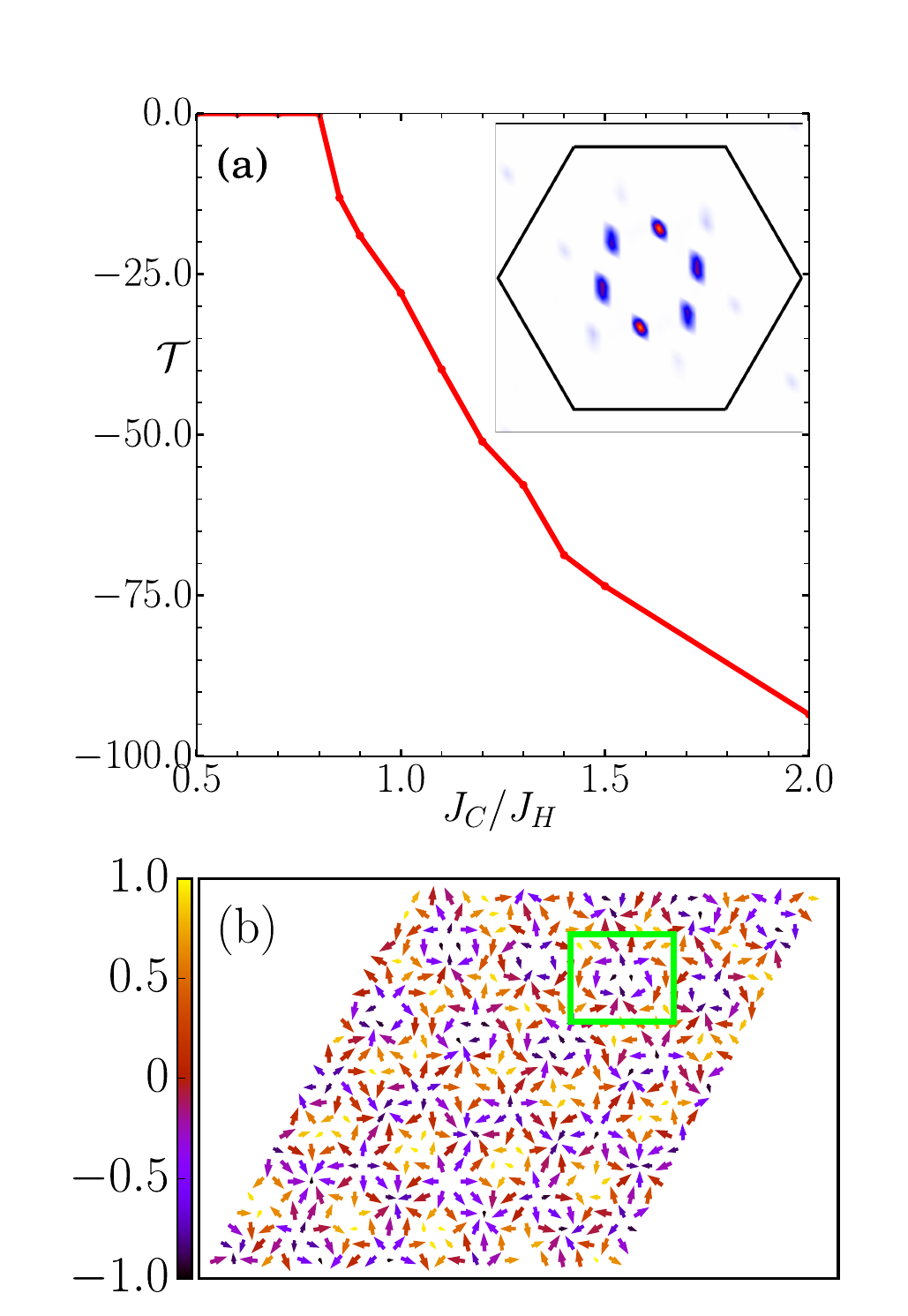}
	\caption{(Color online) (a) Skyrmion density, $\mathcal{T}$ vs $J_C/J_H$ plot with $J_H = -1.0$ fixed. (b) Magnetic ground state at $J_C = -1.5$. Inset in (a) shows the spin structure factor plot coreesponding to spin configuration in (b).
	}
	\label{fig:skyden_vs_Jc}
\end{figure}

Next we explore how the strength of chiral magnetic interaction affects the stabilization of biskyrmions. In the (\cref{fig:skyden_vs_Jc}($a$)), the variation of skyrmion density, $\mathcal{T}$ with $J_C/J_H$ is shown. Varying $J_C$ with $J_H = -1.0$ is taken for this calculation. We can see that below a critical value of $J_C$ that is $0.8$, the ground state is ferromagnetic with $\mathcal{T} = 0.0$. Above this critical value we get the biskyrmions and as the $J_C/J_H$ increases $\mathcal{T}$ also increases. This happens because of the increase in the number of biskyrmions as shown in (\cref{fig:skyden_vs_Jc}($b$)) for the case of $J_C/J_H = 1.5$. We can also see that with increasing strength of $J_C/J_H$, the size of the biskyrmions decreases. This decrease in size of bisyrmions also reflects in the corresponding spin structure factor plot shown in inset of (\cref{fig:skyden_vs_Jc}($a$)). The hexagonal peaks in the spin structure factor occur at large {\bf Q} as compared to the peaks in (see \cref{fig:spin}($b$ and $e$)).

\noindent



\noindent
\underline{\it Conclusion:}  We have presented a new mechanism for stabilizing biskyrmions by scalar chiral spin interaction. Such interactions arise as effective spin interaction in the low energy description of many Mott insulators with strong electron-electron interaction in the presence of an external magnetic field. Using large scale Monte Carlo simulations, we demonstrate the emergence of bisyrmions on a triangular lattice with competing nearest neighbor ferromagentic Heisenberg exchange and a chiral spin interaction defined on the triangular plaquettes. The model can be generalized to other two dimensional lattices. We identify the biskyrmions by their real space spin configuration maps and the corresponding hexagonal peaks in the static spin structure factor plots. We further characterize the nature of these biskyrmions by their skyrmion densities. The number and the size of the biskyrmions are shown to depend on the strength of the chiral spin interaction parameter.

\noindent
\underline{\it Acknowledgments:} 
We acknowledge financial support from the Ministry of Education, Singapore through grant RG 138/22 and the use of the computational resources at the High Performance Computing Centre (HPCC) at NTU, Singapore and the National Supercomputing Centre (NSCC), Singapore.

\end{document}